\RequirePackage{fix-cm}
\documentclass[smallextended]{svjour3}       
\smartqed  
\usepackage{amsmath,amsfonts,amssymb}
\usepackage[latin1]{inputenc}
\usepackage{graphics} 
\usepackage[colorlinks=true, allcolors=blue]{hyperref} 
\usepackage[ngerman,english]{babel} 
\usepackage{lineno}
\usepackage{lscape} 
\usepackage{rotating} 
\usepackage{color}
\usepackage{soul}
\usepackage{xcolor,colortbl}
\usepackage{marvosym}
\newcommand{\envelope}{(\raisebox{-.5pt}{\scalebox{1.45}{\Letter}}\kern-1.7pt)}

\begin{document}

\title{Procedures for the relative calibration of the SiPM gain on ASTRI SST-2M camera}


\titlerunning{Procedures for the relative calibration of the SiPM gain}        

\author{D. Impiombato,          
        O. Catalano,\\          
		S. Giarrusso,          
		T. Mineo,              
		G. La Rosa,\\            
		C. Gargano,            
		P. Sangiorgi,          
		A. Segreto,\\            
		G. Sottile,            
		G. Bonanno,            
		S. Garozzo,\\            
		A. Grillo,             
		D. Marano,             
		G. Romeo,\\
		R. Gimenes
		}

\authorrunning{D. Impiombato, O. Catalano, S. Giarrusso, T. Mineo and G. La Rosa et al.} 

\institute{D. Impiombato\envelope \at 
              INAF, Istituto di Astrofisica Spaziale e Fisica cosmica di Palermo, via U. La Malfa 153, I-90146 Palermo, Italy  Tel.: +39 0916809468 Fax: +39 0916882258\\
			  \email{Domenico.Impiombato@iasf-palermo.inaf.it} \\
           O. Catalano, S. Giarrusso, T. Mineo, G. La Rosa, C. Gargano, P. Sangiorgi, A. Segreto, G. Sottile \at
              INAF, Istituto di Astrofisica Spaziale e Fisica cosmica di Palermo, via U. La Malfa 153, I-90146 Palermo, Italy \\ 
		   G. Bonanno, S. Garozzo, A. Grillo, D. Marano, G. Romeo \at
              INAF, Osservatorio Astrofisico di Catania, via S. Sofia 78, I-95123 Catania, Italy \\
			  R. Gimenes \at
              Instituto de Astronomia, Geofísica e Ciências Atmosféricas, Universidade de São Paulo Rua do Matão 1226, 05508-090 São Paulo, SP, Brazil \\
			}

\date{Received: date / Accepted: 11/11/2016}

\maketitle

\begin{abstract}
ASTRI SST-2M is one of the prototypes of the small size class 
of telescopes proposed for the Cherenkov Telescope Array.  
Its optical design is based on a dual-mirror Schwarzschild-Couder
configuration, and the camera is composed by a matrix of 
monolithic multipixel silicon photomultipliers managed by 
ad-hoc tailored front-end electronics.
This paper describes the procedures for the gain calibration 
on the ASTRI SST-2M.  
Since the SiPM gain depends on the operative voltage and the 
temperature, we adjust the operative voltages for all sensors to 
have equal gains at a reference temperature.
We then correct gain variations caused by temperature changes by 
adjusting the operating voltage of each sensor. 
For that purpose the SiPM gain dependence on operating voltage and 
on temperature have been measured.
In addition, we present the calibration procedures and the results
of the experimental measurements
to evaluate, for each pixel, the parameters necessary to make  
the trigger uniform over the whole focal plane.

\keywords{Front-End; ASIC for SiPM; ASTRI; CTA}

\end{abstract}

\section{Introduction}
The use of Imaging Atmospheric Cherenkov Telescopes (IACTs) in gamma-ray
 astronomy opened a new 
frontier in the study of the emission from high-energy sources such 
as supernovae, neutron stars and supermassive black-holes.
Since the discovery of TeV emission from the Crab nebula with the Whipple 
10m \cite{weeks89} in 1989, the number of known sources emitting in this range 
of energy has rapidly increased to hundreds in a relatively small number 
of years.
The detection of Very High Energy (VHE) gamma photons is possible by observing the 
Cherenkov light produced by the relativistic secondary particles
in the air shower generated by the interaction with the Earth's atmosphere. 
 This light, emitted within a cone of $\sim$1.3$^\circ$ in the ultraviolet band,  
reaches the ground in a ''pool'' with $\sim$120 m radius.
 It is very faint and lasts only a few ns with a duration that depends 
mainly on the distance between the shower core and the telescope 
axis \cite{heb11}.

Significant progress on the investigation of  VHE gamma-ray 
phenomena has been achieved in the last years with the development 
of more complex  IACT instruments such as H.E.S.S. \cite{hoffman99}, MAGIC \cite{ferenc05} 
and VERITAS \cite{holder08} which represent the current state-of-the-art. 
Nevertheless, lots of puzzles still remain in the understanding 
of different features of the TeV gamma-ray astronomy. 
So, the international TeV astrophysics community is moving 
towards a newer generation of IACT  telescopes that
will form the Cherenkov Telescope Array (CTA) observatory \cite{acharya13}. 

CTA will be the largest instrument for gamma-ray astronomy ever conceived 
and will comprise all advantages of the IACT technique. 
The CTA observatory will consist of an extended array of more than  
one hundred of telescopes of different sizes (large, medium, small) 
deployed over two sites, one in each hemisphere.
 At its southern site, CTA plans to install about 70 small size 
 telescopes (SSTs), dedicated  to study the sky from a few TeV up 
 to hundreds of TeV. 
The deployment of the telescopes will be done at different times, 
along intermediate steps. At the beginning, during the so-called 
CTA pre-production phase, small subset of telescopes will be 
installed to assess readiness for mass production. 
At the CTA Southern site, during the pre-production phase, 
it is foreseen the deployment of the ASTRI mini-array led by Italian National 
Institute of Astrophysics (INAF) in synergy with the Universidade 
de S\~ao Paulo (Brazil) and the North-West University (South Africa). 
The ASTRI mini-array \cite{lapalombara14,vercellone15}
will be composed of nine dual-mirror small size telescopes whose end-to-end 
prototype, named ASTRI SST-2M \cite{astri13} and developed by INAF, has
been recently inaugurated in Italy. 
Several challenging but innovative technological solutions characterize
the ASTRI concept \cite{pareschi13}:
the telescope adopts a dual-mirror (2M) Schwarzschild-Couder optical design \cite{canestrari13} 
and the camera at the focal plane is composed of a matrix of Silicon 
photomultipliers (SiPMs) managed  by an appropriate front-end electronics \cite{catalano13}.
The telescope design is based on a focal ratio of 0.5, 
leading to an equivalent focal length of 2.15 m and 
an average effective collecting area of about 6.5 m$^2$ 
with a resulting full field of view, FoV,  
of $\sim$9.6$^\circ$.

This paper describes the procedures and the lab measurements 
necessary for the ASTRI SST-2M SiPMs gain calibration. 
The main purpose of these procedures is to keep during
the normal data taking the gain uniform 
and stable without the aid of  instrumentation external to the camera, correcting for 
variation compared to the nominal values of the involved parameters
 (relative calibration).
These objectives are reached providing the camera with a set of 
reference values for operative voltage and with some coefficients 
to correct gain variations.
Measurements aimed to determine the values of the gain variation as 
a function of the operative voltage and temperature are presented in this paper 
together with the level of accuracy due to different sources of errors.

In addition, we present the calibration procedures 
and the results of measurements performed in lab
to evaluate, for each pixel, the parameters for making uniform 
the trigger over the  whole focal plane.
 
 \section{ASTRI Camera}
 \label{camera}
\begin{figure*}[h!]
\centering
\includegraphics[angle=0, width=10cm]{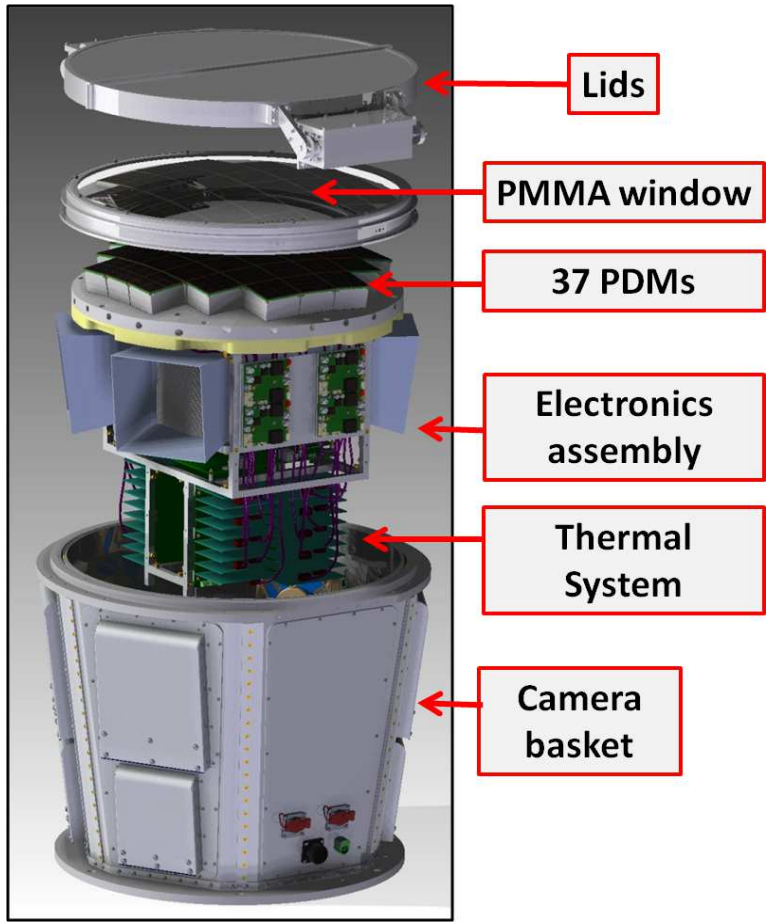}
\vspace{0.5cm}
\caption{Exploded view of the Camera body assembly.}
\label{fig1}
\end{figure*}

The ASTRI SST-2M camera
has a truncated-cone shape and its overall dimensions are
665mm$\times$665mm$\times$515mm, including mechanics and
interfaces with the telescope structure, for a total weight
of $\sim$70 kg. 
An exploded 3D visualization of the camera assembly is 
shown in Figure~\ref{fig1}. 
The focal surface is composed by a matrix of 
SiPM Hamamatsu Silicon
Photomultiplier S11828-3344M \cite{hamamatsu11} working at an operative
voltage of $\sim$70 V.
Each SiPM sensor is formed by 4$\times$4 squared
physical pixels, 3mm$\times$3mm, made up of 3600 
single-photon avalanche diode (SPAD) of
50 $\mu$m pitch  \cite{marano14}.

\begin{figure*}[h!]
\centering
\includegraphics[angle=0, width=12cm]{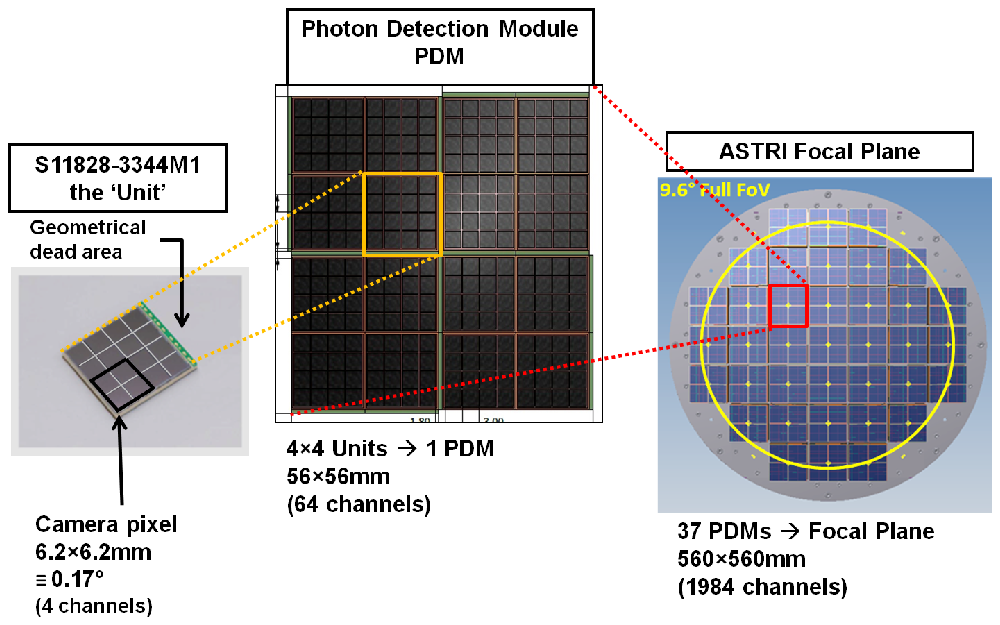}
\vspace{0.5cm}
\caption{
The black square in the left figure shows one of the camera pixels 
 obtained grouping 2$\times$2 physical pixels. The central figure shows
 one ASTRI SST-2M PDM composed of 4$\times$4 SiPMs. The right figure
 represents how the PDMs are arranged on the Focal Surface Camera}

\label{fig2}
\end{figure*}

In order to match
the angular resolution of the optical system,
the physical pixels of each sensor unit are grouped 2$\times$2 
in the so-called camera pixel (6.2mm$\times$6.2mm) having a sky-projected
angular size of 0.17$^\circ$. 
The aggregation of 4$\times$4 sensor units
(8$\times$8 camera pixels) forms a Photon Detection Module
(PDM), and 37 PDMs
form the camera (see Figure~\ref{fig2}).
This modular design allows to better follow the curvature of the 
focal plane and makes easier the maintenance of small portions of the 
camera being the PDMs physically independent.

In order to protect the camera sensors from the external
atmospheric environment, an optical-UV transparent
Poly Methyl MethAcrylate (PMMA)  window 
is mounted onto the focal surface support structure covering all the
PDMs.
This window is modeled with the same radius of
curvature of the focal surface (see Sect.~\ref{focal plane}).
A side emitting Fiber Optics system(FOC) is located  along the PMMA 
circumference. It is illuminated by a Light Emitting Diode (LED) 
system and used for the on-field relative calibration of the gain. 
Its light uniformly illuminates the PDMs units at a controlled intensity.

The ASTRI SST-2M camera
is equipped also with a light-tight lid composed of two petals
mounted onto the backbone structure of the camera. 
This prevents accidental sunlight exposure of the focal surface
detectors, catastrophic in case of direct light reflected
by the mirrors.  

The camera is thermally controlled to keep the
working temperature on the focal plane within the range 13-17$^\circ$C
giving a gain accuracy compliant with the CTA 
calibration requirements
of $\sim$10\% \cite{cta11}.
The temperature of each PDM  is monitored by a system of sensors 
embedded in the SiPM board.

\subsection{Focal plane} 
\label{focal plane}
The Focal Surface Camera (FSC), as shown in Figure~\ref{fig1}, has a 
convex-shaped structure with a curvature radius of 1060 mm.
PDMs, that are square flat modules, are symmetrically placed
on the FSC with angles respect to the telescope axis 
opportunely chosen in order to follow the curvature of the focal plane.
Each PDM, composed of 4$\times$4 side-by-side SiPMs, 
is installed on a Printed Circuit Board (PCB) 
of the same dimension.
The PCB hosts the Front-End Electronics (FEE) composed of
 two ASICs for the signal read out,
four Analogue-to-Digital Converters (ADC) and one Field Programmable
Gate Array (FPGA). This FEE-FPGA manages the ASICs slow control,
the generation of local trigger and the data analog-to-digital conversion.
The Back-End Electronics (BEE) is composed of two main parts that are
a Processing System (PS) and a Programmable Logic (PL). 
It is the main elaboration unit of the camera which
controls and manages the overall system, including data, and
all ancillaries used to perform operations as
the camera thermal regulation, 
the Voltage Distribution management and the time events stamping.
The BEE also provides the functions necessary to
process and transmit the event data as 
obtained by the FEE to an external Data Acquisition 
Workstation responsible for receiving and storing
the data packets.

\subsection{Signal processing}
\label{Sigpro}

\begin{figure*}[h!!]
\centering
\includegraphics[angle=0, width=13cm]{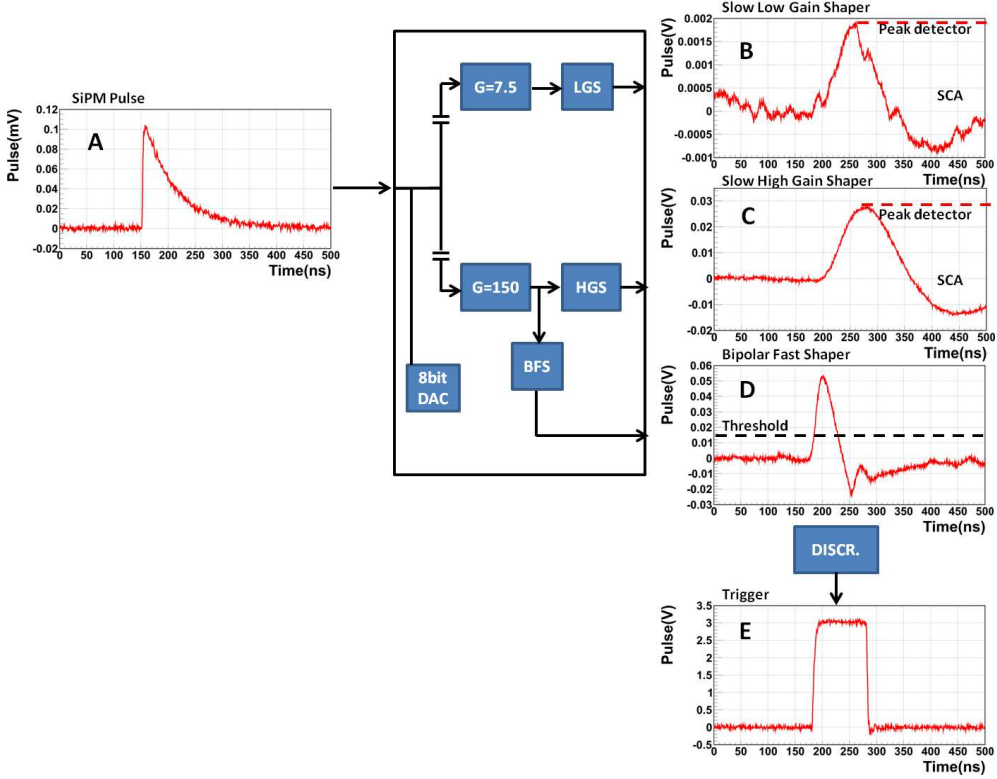}
\vspace{0.5cm}
\caption{Signal processing path of one CITIROC channel.}
\label{fig3}
\end{figure*}

A dedicated FEE has been designed to catch the very fast pulses 
of Cherenkov light and also to provide auto-trigger
capability.
The electronics adopted for the ASTRI SST-2M telescope prototype is 
the Cherenkov Imaging Telescope Integrated Read Out Chip 
(CITIROC \cite{fleury14}), INAF design intellectual property.
The CITIROC chip is equipped with 32 channels capable of measuring charges 
from 1 to 2000 photoelectrons, assuming a SiPM gain of 10$^6$.
A detailed characterization of the front-end electronics
is given in \cite{impiombato15}. 
Two CITIROC devices are devoted to process the SiPM detector signals 
of each single PDM.
Figure~\ref{fig3} represents a schematic view of the whole signal 
processing path in any CITIROC channel.

The wide dynamic range necessary for the camera read-out is provided
by CITIROC using two separated chains integrated 
in the chip: the High Gain (HG) and the Low Gain (LG).

In any CITIROC channel an 8-bit digital 
to analogue converter (DAC) is integrated, by 
varying the operative voltage of the corresponding camera pixel.
The pulse generated by the SiPM (curve A) is amplified by the two 
 programmable 
 preamplifiers (HG and LG) and then shaped with a constant shaping time 
 of 37.5 ns by two electronic circuits (HG and LG).
Two operation modes are implemented
in the HG and LG chains of the ASICs (curves B and C in Figure ~\ref{fig3}): 
the peak detector and the Switched Capacitor Array (SCA).
The peak detector mode, that is the base-line operating approach, provides
the maximum of the shaped pulse height,
whereas the SCA mode returns the level of 
the shaped pulse height 
at a sampling time delayed from the trigger by
a fixed interval.
This delay, that depends on the shaping time, 
is set in order to get the maximum slow shaper signal. 
In both modes, the conversion of the height of the shaped pulse
to the integrated charge introduces a systematic error lower than 1\%.

 A Bipolar Fast Shaper (BFS) 
 connected to the high gain preamplifier  produces 
 a fast signal (curve D) with a shaping time of $\sim$15 ns.  
 It is connected to a 
 discriminator with a 10 bit programmable threshold. 
 The output of the discriminator gives a  digital trigger (curve E) 
 if the signal exceeds the threshold level. 
Digital triggers are available on the 32 + 32 output pads of the 
two ASICs. 
A local trigger is activated when $n$ contiguous pixels
within a PDM  present a signal above the discriminator threshold.
This signal is then propagated to all PDMs to start 
the acquisition of an event in all the focal plane.  
The number $n$ of contiguous pixels, programmable from 2 to 7 and
 the trigger threshold, also programmable ($\geq$4pe) depending 
 on the Night Sky Background (NSB), 
 are chosen in order to have a maximum rate of $\sim$600 Hz  from the 
whole camera, with a dead time $<$3\%,
well inside the maximum fraction of the required dead time 
by CTA \cite{actis11}.

 The signal outputs are read and 
 converted to digital counts by external 
 ADC devices, at the occurrence of a trigger condition.
 A Xilinx Artix 7 FPGA governs and controls all the input/output operations
 from/to CITIROCs and SiPMs.

 \section{Relative calibration plan}

The procedure for the relative calibration of the gain adopted in 
ASTRI SST-2M camera foresees different steps:
\begin{itemize}
\item Define the nominal values of the configuration parameters
 as  temperatures,  gains,  operative voltages, etc.;
\item Determine the parameters of the gain linear variations with
 operative voltage and  temperature;
\item Evaluate the level of accuracy 
for the gain.
\end{itemize}

When the camera is mounted on the telescope, 
these results are used to check and correct the gain relatively
to the nominal values.
These corrections are performed in real time,
taking into account the measurements of the embedded thermal sensors.

In addition, further measurements are periodically 
carried out using the FOC
whose data are used during normal data analysis for off-line corrections of 
the gain.  

All the calibration parameters and procedures 
will be stored and implemented in the BEE.

\subsection{Gain of SiPM camera pixel }
\label{Gain}
The SiPM output signal is produced as a result of an avalanche 
multiplication process characterized by the 
gain $G$, defined as the number of electrons 
 generated in response to a photon absorption or a thermally generated 
 ignition in the silicon lattice. In Geiger-mode operation, the 
 multiplication factor of an avalanche discharge is expected to 
 grow linearly with the operative voltage according to:
 \begin{equation}
\label{fm1}
 G\,=\,\frac{Q_{TOT}}{q}= \frac{C_{pixel}(V_{op}-V_{BD})}{q}
\end{equation} 
\noindent
where $Q_{TOT}$ is the total charge generated by a single avalanche,
 $C_{pixel}$ is the overall pixel capacitance, $q$ is the 
electron charge, $V_{op}$ is the operative voltage and $V_{BD}$ is the 
 detector breakdown voltage.
 
The gain is a function of 
the differences between the two voltages $V_{BD}$ and $V_{op}$.
The former is   
a characteristic of the SiPM physical pixel and is mainly sensitive to
temperature; the latter is different in each camera pixel and can be
opportunely varied to keep the gain uniform over the camera surface 
and stable in time against temperature variations in $V_{BD}$.
In our case, grouping four physical pixels in one camera pixel
 and applying  a single operative voltage
introduces an intrinsic spread in the gain.

Any gain variation $\Delta G(V_{op},T)$ can be written in terms of 
variations of the operative voltage $\Delta V_{op}$ and 
of the temperature $\Delta T$ 
with the following equation:

\begin{equation}
\label{fm2}
 \Delta G(V_{op},T)\,=\,\frac{\partial G}{\partial V_{op}}\Delta V_{op}+\frac{\partial G}{\partial T}\Delta T
\end{equation} 
\noindent
where $\partial G/\partial V$ is the gain derivative with respect to 
the voltage variation, and $\partial G/\partial T$ 
is the gain derivative with respect to the temperature. 

For small variations of the temperature and of the operative voltage, 
$V_{op}$, $\partial G/\partial V$ and $\partial G/\partial T$ can be 
considered constant in a first approximation, and the gain can be computed 
through the following formula:

\begin{equation}
\label{fm3}
 G(V,T)\,=\,c_{1}\cdot \Delta V_{op}+c_{2}\cdot \Delta T+c_{3}
\end{equation} 
\noindent
where $c_{1}=\partial G/\partial V$, $c_{2}=\partial G/\partial T$,
and $ c_{3}=G(V_{0},T_{0})$ is the 
reference gain at a given operative voltage and temperature.
The reference gain suggested by manufacturer is 7.5$\times10^{5}$.
However, we assume $c_{3}$ =6.0$\times10^{5}$ as the reference gain
for all camera pixels 
and  $T_{0}$ =15$^\circ$C as the reference temperature. 
Our choice is mainly driven by the need to reduce the cross talk 
to a value of 12\%.
 
The reference operative voltage $V_{0}$ is set for each camera pixel 
in order to obtain the reference gain.
The average value of  $V_{op}-V_{BD}$ (overvoltage) 
is $\sim$1.5 V.

To keep $G(V,T)$ constant we have to compensate for the temperature  
variations by changing $V_{op}$
according to the relation: 
 \begin{equation}
\label{fm4}
\Delta V_{op}= -\frac{c_{2}}{c_{1}}\cdot \Delta T
\end{equation} 
\noindent

\subsection{CITIROC DAC calibration}

The operative voltage $V_{op}$ applied to each camera pixel is 
the sum of the external voltage $V_{ext}$ and the voltage applied
from the input programmable DAC implemented 
in the CITIROC front-end, $V_{DAC}$, whose dynamical range is $\sim$ 4.5 V:  
\begin{equation}
\label{fm7}
V_{op}\,=\,V_{ext}-V_{DAC} 
\end{equation} 
\noindent
In the CITIROC ASIC, the $V_{DAC}$ is linearly dependent on the DAC code
according to the following equation:

\begin{equation}
\label{fm8}
V_{DAC}\,=\,a\cdot DAC+b 
\end{equation} 
\noindent
where $a$ and $b$ are characteristic of each channel, and then at variance with
$c_{1}$ and $c_{2}$ must be measured for all camera pixels.

$V_{DAC}$ can be used to 
adjust the SiPM gain, compensating for temperature variations.
Combining equations (\ref{fm7}) and (\ref{fm8}), the formula in (\ref{fm4})
can be written in terms of $V_{DAC}$ to obtain the following relation
implemented in the BEE-FPGA:  
\begin{equation}
\label{fm9}
\Delta DAC \,=\,-\frac{c_{2}}{c_{1}}\cdot \frac{\Delta T}{a}
\end{equation} 
\noindent

\subsection{Temperature}
The calibration of the temperature sensors embedded in 
each SiPM board is an 
essential prerequisite for the SiPM gain equalization. 
Each SiPM PCB has 9 temperature sensors symmetrically 
spaced on the rear side of the board.
They operate in the range
from -40$^\circ$C to +125$^\circ$C with a precision of $\pm$0.2$^\circ$C.
These sensors are constantly monitored by the FEE-FPGA, 
and the temperatures measured in each PDM  are sent to the BEE
in order to adjust the gain of the pixels.
The sensors are read at a rate 
of few Hz, adequate for the expected slow-time temperature 
variations inside the camera.
Only the central sensor at 
the bottom of the board has been absolutely calibrated using a
temperature sensor and an appropriate controller provided by LakeShore\footnote{http://www.lakeshore.com/Documents/LSTC\textunderscore 325\textunderscore l.pdf}.
The calibration of the other eight sensors relatively to the central one
will be accomplished in lab during the functional tests of each PDM.

\section{Laboratory Experimental Set-Up}

\begin{figure*}[h!!]
\centering
\includegraphics[angle=0, width=11cm]{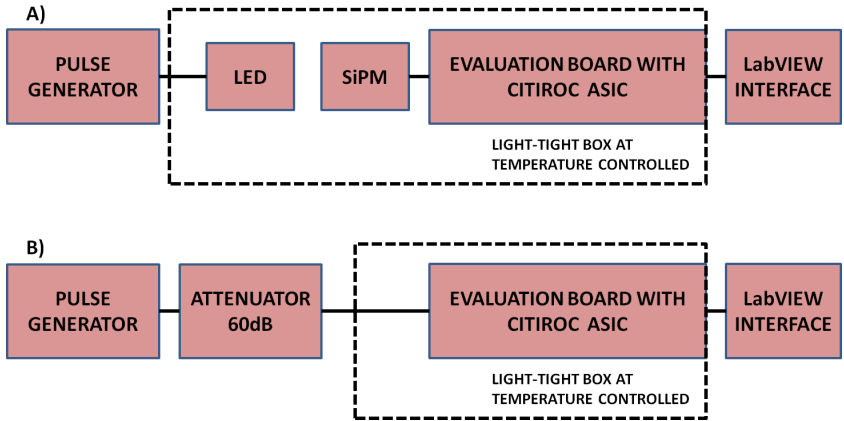}
\vspace{0.5cm}
\caption{Experimental set-up. A light-tight box coupled with a Peltier
 Cooling/Heating system assures a constant temperature on the SiPM 
 under test with stability of $\sim$ 0.1$^\circ$C.}
\label{fig4}
\end{figure*}

All lab measurements have been carried out in the laboratory
of IASF-Palermo and OACT-Catania.
For convenience, all measurements were performed 
using an evaluation board containing one CITIROC.
Two different sets of measurements were performed: 
\begin {itemize}
\item the SiPM pixels were illuminated by a Blue Light Diode (B-LED) 
driven by a pulse generator at a constant rate of 10 kHz.
The schematic overview is shown in Section $A$ of Figure~\ref{fig4}.
This set-up was used for the $c_{1}$ and $c_{2}$ 
evaluation. 
The gain relative to each measurements was obtained
from the average distance between the peaks in the Pulse Height
Distributions (PHDs) as explained later on, in Sect.~\ref{coefficients}.
The value of $c_{1}$ is obtained 
keeping the temperature constant and varying $V_{op}$ 
 in the operating range in steps of 100 mV.
The coefficient $c_{2}$ is measured keeping 
the operative voltage to the reference value $V_{0}$ and varying the temperature in 
 steps of about 1 $^\circ$C from 13$^\circ$C to 17$^\circ$C.
\item An arbitrary  
generator was used to create an waveform signal input signal as similar as possible 
to the SiPM one (rise time of a few ns, recovery time of 175 ns and 1pe 
amplitude).The input signal was 
injected in the channels of the CITIROC evaluation board, 
as shown in in Section $B$ of the Figure~\ref{fig4}.
These measurements were used to calculate 
the ADC photoelectron conversion factor 
 from the obtained PHDs (see Sect.~\ref{coefficients}).
\item a Keithley
 high impedance multimeter (not shown Figure~\ref{fig4})
was used to evaluate the coefficients $a$ and $b$ of the linear relation 
DAC voltage-DAC codes (see Sect.~\ref{ab}). 
 
\end {itemize}

The laboratory equipment includes 
a signal generator Agilent 81160A plus a precise attenuator to be used
with the pulse signal.
The CITIROC evaluation board is driven by a  
Lab-VIEW software\footnote{http://www.ni.com/labview/i/} interface
 provided by Weeroc\footnote{http://www.weeroc.com/}.

\section{Results}

 In the following subsections, results from lab measurements relative to
a single camera pixel, taken as example, are reported.

\subsection{Calibration of the reference voltage $V_{0}$}
\label{calibration}

\begin{figure*}[h!!]
\centering
\includegraphics[angle=0, width=10cm]{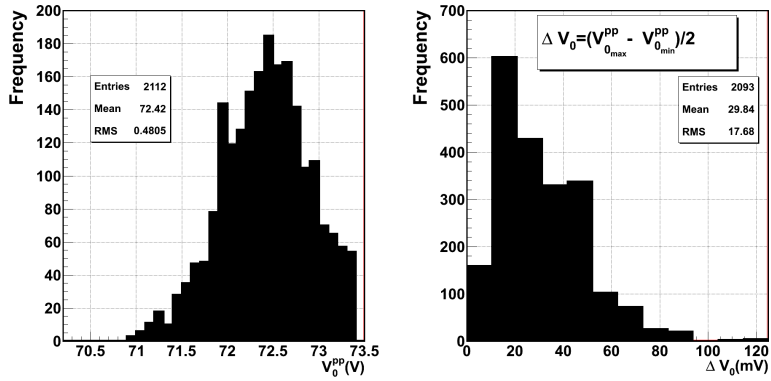}
\vspace{0.5cm}
\caption{Histogram of the measured $V_{0}^{pp}$ of all physical pixels 
(left panel) and of the $\Delta V_{0}$ of the camera pixels (right panel).}
\label{fig5}
\end{figure*}

The reference values of the operative voltage $V_{0}^{pp}$
for each physical pixel, for a gain of 
7.5$\times$10$^{5}$ at the temperature of 25$^\circ$C, were provided
by Hamamatsu.
These values were verified on a sample in our lab.
The values are distributed in the range 71--73.5V as shown
in the histograms of the left panel of Figure~\ref{fig5}.
However, grouping four physical pixels in one camera pixel and adopting
a single operative voltage $V_{0}$ introduces a spread in the gain.
This was evaluated measuring the difference ($\Delta V_{0}$)
between the minimum and the maximum $V_{0}^{pp}$  of the four physical pixels.
The histogram of this spread is shown in the right panel 
of Figure~\ref{fig5}.
 
This average gain uncertainty  
can be computed using the mean value
of $\Delta V_{0}$ (29.84 mV) and the parameter $c_{1}$, 
presented in Sect.~\ref{coefficients}.
This corresponds to a gain 
variation of 1.5\% of the Hamamatsu nominal 
gain (7.5$\times10^{5}$).

\newpage
\subsection{Determination of $a$ and $b$ coefficients}
\label{ab}
\begin{figure*}[h!!]
\centering
\includegraphics[angle=0, width=10cm]{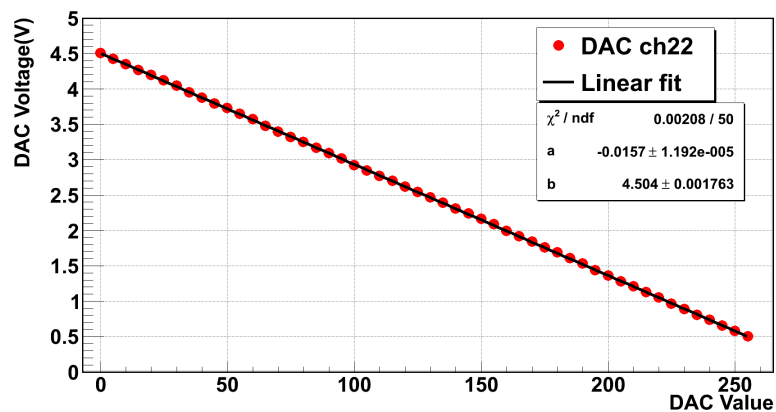}
\vspace{0.5cm}
\caption{CITIROC evaluation board DAC voltage as a 
function of the DAC code for one of the channels.
The error of each point is less than 1\%, so it is not visible 
in the figure.}
\label{fig6}
\end{figure*}

The coefficients of the linear fit of the DAC voltage as a 
function of the DAC codes are obtained 
 reading the output voltage of the DAC by a Keithley
 high impedance multimeter, and varying progressively the DAC codes.
 Figure~\ref{fig6} shows the DAC voltage   
 as a function of the DAC codes for one of the channels.
 
\begin{figure*}[h!!]
\centering
\includegraphics[angle=0, width=10cm]{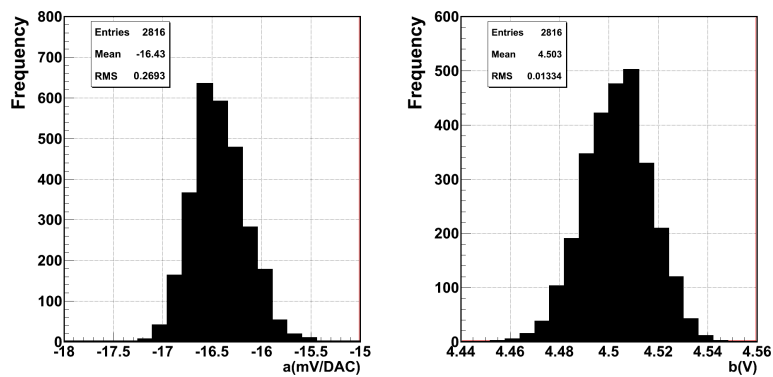}
\vspace{0.5cm}
\caption{Histograms of $a$ and $b$ coefficients for all PDM}
\label{fig7}
\end{figure*}

Each channel has a slightly different slope, 
so this measurement is required for all camera pixels.
 The histograms of $a$ and $b$ coefficients 
are shown in Figure~\ref{fig7} in the left 
and right panel, respectively.

The values of $V_{op}$ can be linearly 
adjusted channel by channel in the interval 0 V$\div$4.5 V. 
This dynamical range is larger than the spread in the values of 
$V_{0}^{pp}$ (2.5 V), as shown in the left panel of Figure~\ref{fig5},
and larger than the corrections needed for compensating temperature variations
(71 mV/$^\circ C$ equivalent to $\sim$4 DAC/$^\circ C$).

The average value of $a$ that is 
 the DAC input resolution,
corresponds to a gain 
variation of 6.32$\times10^{3}$ that is  1.1\% of the nominal 
gain.

 \subsection{ Determination of $c_{1}$ and $c_{2}$ coefficients }
 \label{coefficients}
\begin{figure*}[h!!]
\centering
\includegraphics[angle=0, width=10cm]{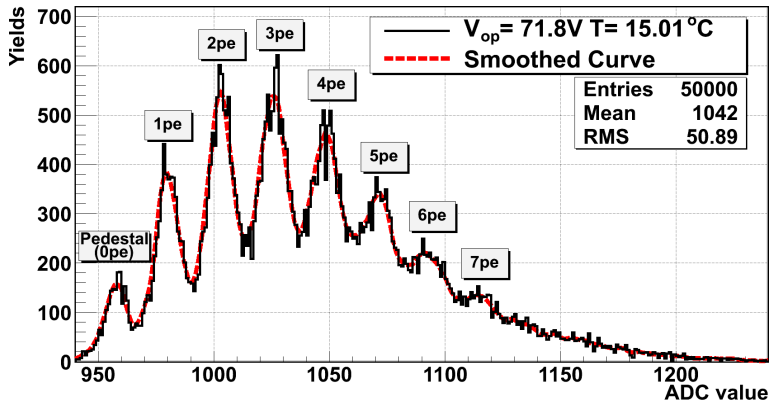}
\vspace{0.5cm}
\caption{CITIROC: PHD at a fixed temperature of 15$^\circ$C
and $V_{0}$=71.8 V for one of the channels.}
\label{fig8}
\end{figure*}

 The coefficients of the gain variations $c_{1}$ and $c_{2}$ 
 are determined with the PHD  pulsing a
 blue LED at a constant rate of 10 kHz and a time-on duration of 11.5 ns.
An example of a PHD at a fixed voltage $V_{0}$= 71.8 V and the
temperature of 15$^\circ$C is shown in Figure~\ref{fig8}.

The ADC photoelectron conversion factor $pe_{eq}$
is evaluated averaging all distances of subsequent local
 maxima of the smoothed PHD collected from each SiPM camera pixel.
 This smoothing is performed 
 with a LOWESS (Locally Weighted Scatterplot Smoothing) method.
The $pe_{eq}$ is then computed applying the following equation:
 \begin{equation}
\label{fm15}
pe_{eq}(ADC)\,=\,\frac{1}{N-2}\cdot (ADC_{N}-ADC_{2}) 
\end{equation} 
\noindent
where $ADC_{N}$ is the ADC value corresponding to the last peak 
and $pe_{eq}$ is the photoelectrons equivalent number 
in ADC units.
Note that the pedestal ($N=1$), whose position 
is biased by the electronic noise, is not considered. 
Its distance from the 1pe peak distribution
has been found significantly different from the average
distances between the other peaks in updated versions of sensor
 \cite{impiombato16}.
We decided to exclude it from computation of the gain
in order to implement a procedure that can be exported to newer SiPMs
even if for the sensor adopted in the present paper the pedestal could 
 be coherently included in the computation.
 
The measured $pe_{eq}$ value is then converted in gain
 comparing it with the value $pe_{ref}$ 
obtained by injecting a charge with a pulse shape similar to the
SiPM one corresponding 
to 1 pe at the reference gain, and applying the following formula:
 
\begin{equation}
\label{fm16}
G(V_{op},T)= \frac{pe_{eq}(ADC)}{pe_{ref}(ADC)}\cdot G(V_{0}, T_{0})
\end{equation} 
\noindent

This method allows a finally accuracy on the gain lower than 1\%
over a 50000 event acquisition.

\begin{figure*}[h!!]
\centering
\includegraphics[angle=0, width=10cm]{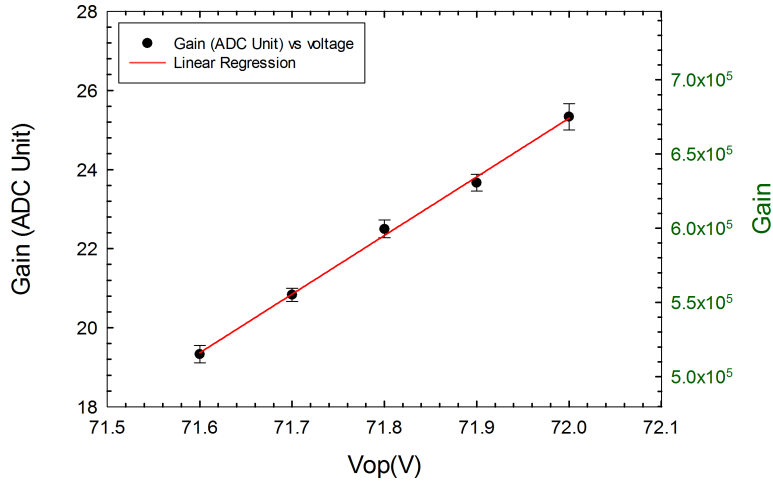}
\vspace{0.5cm}
\caption{Determination of the gain variation coefficient in ADC units as
 a function of $V_{op}$. The ordinate axis on the right 
 reports the estimated pixel gain as a function of $V_{op}$.}
\label{fig9}
\end{figure*}

We measured the values of $c_{1}$ for one of the camera pixel
obtained from PHDs keeping the temperature constant and varying $V_{op}$ 
 in the operating range in steps of 100 mV.
The measured gain is then fitted to a linear function of the operating
voltage, in which $c_{1}$ is the slope of the best fit line.  
In Figure~\ref{fig9}, an example of a gain plot versus $V_{op}$  
for a camera pixel is shown.
The range of the operative voltage 
is 71.60-72V, and the gain increases with voltage with the coefficient
$c_{1}$ equal to 14.76$\pm$0.78 (ADC/V).

\begin{figure*}[h!!]
\centering
\includegraphics[angle=0, width=10cm]{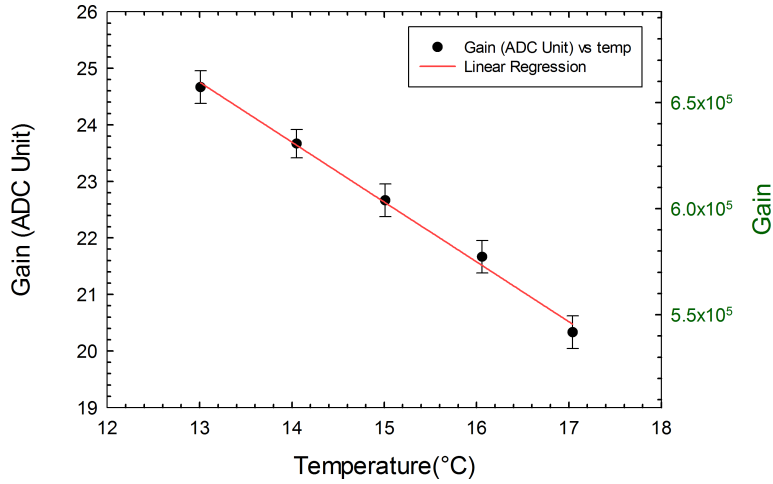}
\vspace{0.5cm}
\caption{Determination of the gain variation coefficient in ADC units 
as a function of temperature. The ordinate axis on the right 
reports the estimated pixel gain as a function of temperature.}
\label{fig10}
\end{figure*}

The coefficient $c_{2}$ is obtained from PHDs  
keeping the operating  voltage to the reference value $V_{0}$ and 
varying the temperature in steps of about 1$^\circ$C 
from 13$^\circ$C to 17$^\circ$C. 
Again, fitting the measured gains to a linear function, 
the value of $c_{2}$ corresponds to the slope.
 In the example plotted in Figure~\ref{fig10}, 
$c_{2}$ is -1.06$\pm$0.09 ($ADC/^\circ C$).
In this case the gain decreases linearly with temperature.

Considering that, in a first approximation, $c_{1}$ and $c_{2}$
are characteristic of the type of SiPM \cite{buhzan01}, we assumed 
a single value for all camera pixels. Discrepancies in the gain
uniformity are corrected off-line using the FOC as already mentioned in
the Sect.~\ref{camera}.

\section{CITIROC trigger calibration}
In addition to the gain calibration we performed in lab the calibration
of the camera pixel trigger.
This has the aim to uniform the trigger response over all camera.
CITIROC is equipped with a 10bit DAC common to all the channels for 
setting the value of each discriminator voltage.
As for any multi-channel electronics front-end, small differences 
 between trigger channels are expected. 
Performing trigger equalization between pixels is essential for a stable
and reliable camera operation.
To compensate the channel-by-channel trigger non-uniformity,
  a 4bit DAC fine-tunes the discriminator threshold level.
 The linear relation between the 10bit DAC and the 4bit DAC is 
 different for each channel. 
The resulting set of 32 4bit DAC codes are stored in the registers 
of each CITIROC providing trigger level discrimination uniformity.

\begin{figure*}[h!!]
\centering
\includegraphics[angle=0, width=10cm]{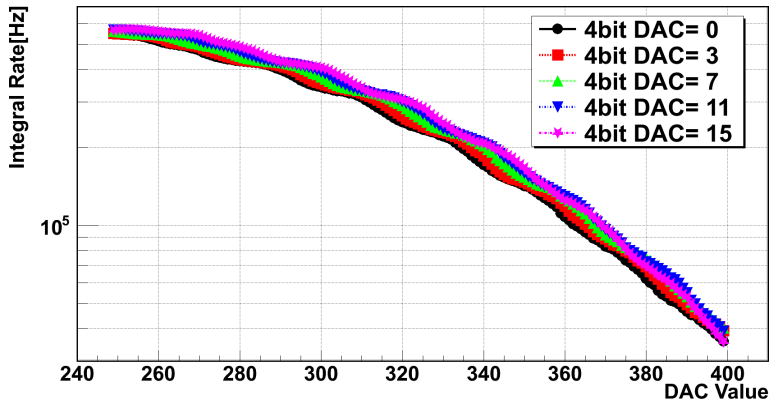}
\vspace{0.5cm}
\caption{ MFS for one of the channels. Trigger rate as a function of the 10bit DAC 
code for five increments of the 4bit DAC (0, 3, 7, 11 and 15).
The error of each point is less than 0.1 \%, so it is not visible 
in the figure.}
\label{fig11}
\end{figure*}

\begin{figure*}[h!!]
\centering
\includegraphics[angle=0, width=10cm]{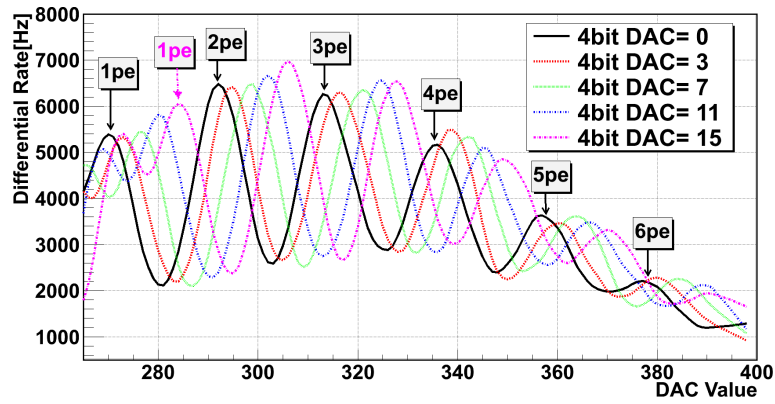}
\vspace{0.5cm}
\caption{Differential trigger rate varying the 4bit DAC in steps of 4
 DAC counts. An increase in the 4bit DAC code results in a shift of 
 the pe peaks in the abscissa.}
\label{fig12}
\end{figure*}

\begin{figure*}[h!!]
\centering
\includegraphics[angle=0, width=10cm]{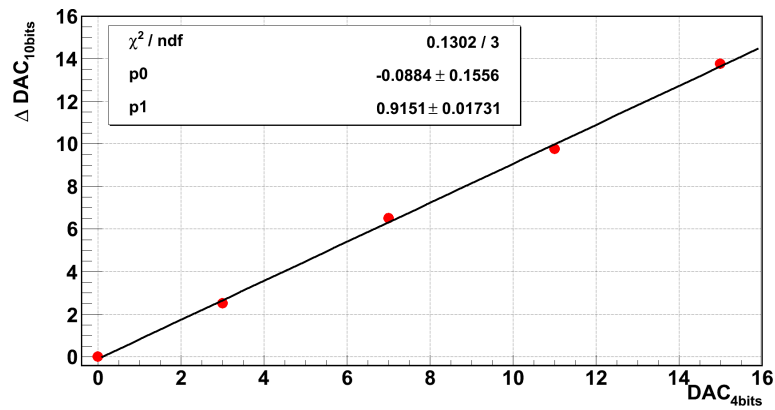}
\vspace{0.5cm}
\caption{4bit DAC linearity. The error of each point is $\sim$0.1 DAC, so 
it is not visible in the figure.}
\label{fig13}
\end{figure*}

The trigger equalization is determined performing Multiple Fiber
 Stair MFS measurements after the gain equalization.
In this case, the SiPM is illuminated with a pulsed LED that produces
a Poissonian distribution with an average number of pe $\sim$4, and
the number of triggers as a function of the discriminator 
threshold level is collected.   
 The envisaged methodology for trigger equalization is developed 
 in several steps as presented below:
\begin{enumerate}
\item
 Perform the MFS measurements, setting to zero the 4bit DAC;
\item Repeat the MFSs, increasing the
 4bit DAC code up to 15;
 Figure~\ref{fig11} shows the trigger rate
as a function of the 10bit DAC code for five different 4bit DAC increments 
in one of the channel.
\item 
  Differentiate the trigger rate curves
 to obtain the distribution of 
 the photons emitted by the LED pulses. 
 Figure~\ref{fig12} shows the differential trigger rate obtained 
  from the curves of Figure~\ref{fig11}.
\item Compute the average distances in DAC units
between adjacent peaks and plot them as a function of the 
4bit DAC code. 
An increase of the 4bit DAC corresponds to a shift of the peak 

\item Fit the data points with the following equation:
\begin{equation}
\label{fm21}
\Delta DAC_{10bits}\,=\,\ m\cdot DAC_{4bits}+q
\end{equation} 
\noindent 
where $\Delta DAC_{10bits}$ is the average displacement in DAC units
between adjacent peaks (see Figure~\ref{fig13}).
It allows to equalize the signals of all trigger discriminators.
\end {enumerate}
 
 The linear relation between the 10bit DAC displacement ($\Delta DAC_{10bits}$) 
 and the 4bit DAC counts for one of the channels is shown in Figure~\ref{fig12}.  

\clearpage
\newpage
\section{Summary and conclusion}

We presented the relative calibration procedures for the 
ASTRI SST-2M camera designed to keep the gain uniform over the focal plane surface
and constant in time. 
In particular, for each pixel, we determined the reference value of the 
operative voltage $V_{0}$ with the relative $DAC_{0}$ at temperature of
 $15^\circ$C and the reference values of the coefficients 
($a$, $b$, $c_{1}$ and $c_{2}$) used to correct, on field, the 
variation of the gain due to temperature variations acting on the 
operative voltage.

The contribution of all
error sources, at one $\sigma$ level, to the total
gain calibration accuracy, 
computed for the reference gain (6.0$\times10^{5}$),
are listed in Table~\ref{table3}.
Note that, the errors introduced by the spreads in $<V_{op}>$
within the camera pixel (see Sect.~\ref{calibration}) are computed 
assuming that the spreads at the gain 6.0$\times10^{5}$ are identical 
to those measured at 7.5$\times10^{5}$. 
This  is coherent with our assumption
that the value of $c_{1}$ is equal for all pixels (see Sect.~\ref{coefficients}).

\begin{table*}[htbp!!]
\centering
\caption{Contributions to the gain error budget in the calibration 
procedure.The first column lists the sources of errors whose level
 is given in the second column; the last column presents the gain error
in percentage for the reference gain 6.0$\times10^{5}$}.
\label{table3}
\begin{tabular}{|l|l|l|}
\hline
                            
Error Source                              & \multicolumn{1}{c|}{Parameter Error}           & \multicolumn{1}{c|}{Percentage}\\ 
\hline                                   
SiPM power supply ripple (mV)             &    $\pm$10                             &  0.7                \\                      
 \hline                                                                                                            
12 bits Pulse Height accuracy (ADC)       &    $\pm$0.1                            &  0.5                \\
 \hline                                                                                                            
Temperature sensors ($^\circ$C)          &    $\pm$0.2                             &  0.9                \\
\hline                                                                             
DAC gain adjustment (8 bits)             &    $\pm$1                               &  1.1                \\                                  
\hline                                                                                                             
SiPM $<V_{op}>$ within pixel             &    1.1$\times10^{4}$                    & 1.8                 \\                                                                                              
\hline                                                                          
$\partial G/\partial V$ (ADC/V)          &    $\pm$0.78                            &  5.3                 \\
\hline                                                                                                                            
$\partial G/\partial T$ (ADC/$^\circ$C)  &    $\pm$0.09                            &  8.3                 \\
\hline
 \end{tabular}
 \end{table*}
\noindent

We note that our total error budget is 10.1\%,  
compliant with the CTA requirements \cite{actis11}.

In addition, we presented the relative calibration of the trigger
necessary to keep uniform the trigger rate on the entire focal plane.

The calibration procedures described in this document 
 are optimized for the current SiPM (Hamamatsu MPPC S11828-3344M), 
 they are still valid for any other SiPM sensors.

\section*{Acknowledgements}
Authors thank the anonymous referee for his/her useful comments that
really improved the paper.
The work presented in this paper was supported in part by the ASTRI,
"Flagship Project" financed by the Italian Ministry of Education, 
University, and Research (MIUR) and  led by the Italian National Institute 
for Astrophysics (INAF). 
We also acknowledge partial support from MIUR Bando PRIN 2009 and 
TeChe.it 2014 Special Grants.







\bibliographystyle{model1c-num-names}
\bibliography{<your-bib-database>}







\end{document}